\documentclass[twocolumn]{article}

\usepackage{lmodern}
\usepackage{authblk}
\usepackage{amsmath}
\usepackage[colorlinks=true, allcolors=cyan]{hyperref}
\usepackage{graphicx,xcolor}

\date{June 15, 2023}

\title{\LARGE \bf Post-compression of multi-mJ picosecond pulses to few-cycles approaching the terawatt regime}

\author[1,2,*,$\dagger$]{Supriya Rajhans}
\author[1,$\dagger$]{Esmerando Escoto}
\author[1]{Nikita Khodakovskiy}
\author[3]{Praveen K. Velpula}
\author[1]{Bonaventura Farace}
\author[1]{Uwe Grosse-Wortmann}
\author[1]{Rob J. Shalloo}
\author[4]{Cord L. Arnold}
\author[1]{Kristjan Põder}
\author[1]{Jens Osterhoff}
\author[1]{Wim P. Leemans}
\author[1]{Ingmar Hartl}
\author[1,5,6]{Christoph M. Heyl}

\affil[1]{Deutsches Elektronen-Synchrotron DESY, Notkestr. 85, 22607 Hamburg, Germany}
\affil[2]{Friedrich-Schiller-Universität Jena, Max-Wien-Platz 1, 07743 Jena, Germany}
\affil[3]{UGC-DAE Consortium for Scientific Research, University Campus, Khandwa Road, Indore 452001, Madhya Pradesh, India}
\affil[4]{Department of Physics, Lund University, PO Box 118, SE-221 00 Lund, Sweden}
\affil[5]{Helmholtz-Institute Jena, Fröbelstieg 3, 07743 Jena, Germany}
\affil[6]{GSI Helmholtzzentrum für Schwerionenforschung GmbH, Planckstraße 1, 64291 Darmstadt, Germany}
\affil[*]{Corresponding author: supriya.rajhans@desy.de}

\begin{document}
\twocolumn[
  \begin{@twocolumnfalse}
    \maketitle

\begin{abstract}
Advancing ultrafast high-repetition-rate lasers to shortest pulse durations comprising only a few optical cycles while pushing their energy into the multi-millijoule regime opens a route towards terawatt-class peak powers at unprecedented average power. 
 We explore this route via efficient post-compression of high-energy 1.2\,ps pulses from an Ytterbium InnoSlab laser to 9.6 fs duration using gas-filled multi-pass cells (MPCs) at a repetition rate of 1\,kHz. Employing dual-stage compression with a second MPC stage supporting a close-to-octave-spanning bandwidth enabled by dispersion-matched dielectric mirrors, a record compression factor of 125 is reached at 70\% overall efficiency, delivering 6.7\,mJ pulses with a peak power of about 0.3\,TW. Moreover, we show that post-compression can improve the temporal contrast at picosecond delay by at least one order of magnitude. 
Our results demonstrate efficient conversion of multi-millijoule picosecond lasers to high-peak-power few-cycle sources, opening up new parameter regimes for laser plasma physics, high energy physics, biomedicine and attosecond science.
\end{abstract}
\vspace{5 mm}
\end{@twocolumnfalse}
  ]

The development of laser sources towards higher energies and shorter pulse duration has shown continuous progress in the last decades and has enabled key applications across multiple fields. One promising direction targets the generation of laser pulses for driving laser-based particle acceleration, which requires peak powers in the terawatt (TW) regime. Most of the demonstrations of laser-based particle acceleration to date have been performed at low repetition rates (typically $\leq$10\,Hz). The technology and prospective applications of compact laser-based accelerators would, however, greatly benefit from repetition rates in the kHz-MHz regime, thus demanding laser sources which provide not only terawatts of peak power but also high average power \cite{Albert_2021_roadmap}. 
Ytterbium (Yb) -based laser sources show better scalability to higher repetition rates and kWs of average power, much beyond widely used Ti:sapphire systems \cite{russbueldt2010compact, russbueldt2009400}. However, the pulse duration achievable with Yb-based amplifiers can only reach a few hundreds of femtoseconds (fs) due to the limited gain bandwidth \cite{jauregui_high-power_2013}. This limitation sets up Yb-based lasers as excellent match with post-compression methods. In particular, the combination of high-energy Yb systems \cite{Innoslab_amplifier} with efficient post-compression methods supporting large compression factors, promises compact, high repetition rate, high peak power few-cycle sources \cite{nagy_high-energy_2021, Viotti_MPC_review}.

Several options for post-compressing Yb lasers exist, mostly relying on spectral broadening through self-phase modulation (SPM). For pulse energies in the few-millijoule range, sufficient to reach the TW peak power range when compressed to the few-cycle regime, hollow-core fibers (HCFs) \cite{jeong_direct_2018, fan_HC_40mJ_compression_2021,Nagy_2019_Yb} and multi-pass cells (MPCs) \cite{Hanna_review_2021,Viotti_MPC_review} have emerged as most commonly employed schemes. A third powerful technique, the multi-plate approach has been employed for compression down to the single-cycle regime, however, the technique reaches limitations at millijoule pulse energies \cite{lu_greater_2019, seo_high-contrast_2020}.  
In contrast, HCFs and MPCs can easily be scaled to higher pulse energies by lengthening the setup, limited mainly by the ionization threshold of the nonlinear medium and the laser-induced damage threshold (LIDT) of the optics. However, higher compression factors can be reached in a more compact setup using an MPC compared to a HCF \cite{Viotti_MPC_review}. 
In addition, gas-filled MPCs have been shown to support pulse energies exceeding 100\,mJ \cite{Kaumanns_2021,Pfaff_100_mJ_2022}, they provide high energy transmission \cite{Grebing_2020,Schulte_2016}, very low sensitivity to beam pointing \cite{vernaleken_patent}, and excellent beam quality \cite{Lavenu_nonlinear_2018}.  
MPCs encounter challenges when approaching the few-cycle regime, in particular in combination with high average power. 
Few-cycle pulses, which require a near octave-spanning spectrum, can be easily supported by metallic mirrors \cite{Rueda_8_2021,balla_postcompression_2020,Muller_multipass_2021}. However, metallic mirrors absorb a significant amount of energy, leading to thermal issues at high average power. 
M\"uller et al. circumvented this issue by using enhanced silver mirrors on water-cooled monocrystalline silicon substrates, demonstrating the compression of 200\,fs pulses to few-cycle duration with a record high average power of 388\,W and a pulse energy of 776\,$\mu$J \cite{Muller_multipass_2021}. 
Alternatively, dielectric mirrors can also support octave-spanning bandwidths if employed as dispersion-matched mirror pairs and have been implemented for few-cycle MPCs \cite{goncharov2023few,viotti2023few}. Furthermore, dielectric mirrors support a higher reflectance (\textgreater99.9\,\%) than metallic mirrors (\textgreater98\,\%) and enable dispersion control \cite{silletti2022dispersion}. However, until today, MPCs using dispersion-matched dielectic mirrors have not entered the millijoule energy regime. 

In this work, we demonstrate high-energy MPC-based post-compression to few-cycle duration using dielectric mirrors. We report record high pulse energies and peak powers achieved by compressing 1.2\,ps, 9.45\,mJ laser pulses at a repetition rate of 1\,kHz to less than 10\,fs duration. The compression scheme has been optimized to support an extreme compression factor exceeding 120 while maintaining an overall throughput of 70\,\% and improved temporal pulse contrast after post-compression. This is achieved employing a two-stage cascaded MPC setup based on all-dielectric mirrors.  

Shortening the pulse duration from 1.2 ps down to few cycles (\textasciitilde10\,fs at 1030 nm) requires a large compression factor of about 120. This is most easily achieved in a multi-stage cascaded compression setup. 
Compared to single-stage approaches \cite{nagy_high-energy_2021}, this relaxes the need for a large amount of B-integral which would compromise the temporal quality of the compressed pulse \cite{Fritsch_beam_quality, Escoto_temporal_quality}. For a Gaussian pulse shape the compression factor can be estimated as $C \approx 1 + 0.59 B$, where $B$ denotes the B-integral \cite{Escoto_temporal_quality}. Thus, more than 200 radians of B-integral are needed to compress picosecond pulses down to a few femtoseconds. However, if the compression is split into two stages, with both stages operating at a similar compression factor, the B-integral per stage is only about 20\,rad. Two compression stages can also contribute to improved peak power and temporal contrast, even when the compression ratios of both stages is not equal \cite{Escoto_temporal_quality}. A non-equal ratio can be more practical and help increasing the overall throughput when considering the bandwidth and LIDT of the mirrors. Since the first stage needs to support a much narrower bandwidth, standard dielectric quarter-wave-stack mirrors with high LIDT can be used, and a larger spectral broadening factor can thus be targeted in the first stage. 

\begin{figure}[!tb]
\centering
\includegraphics[width=\linewidth]{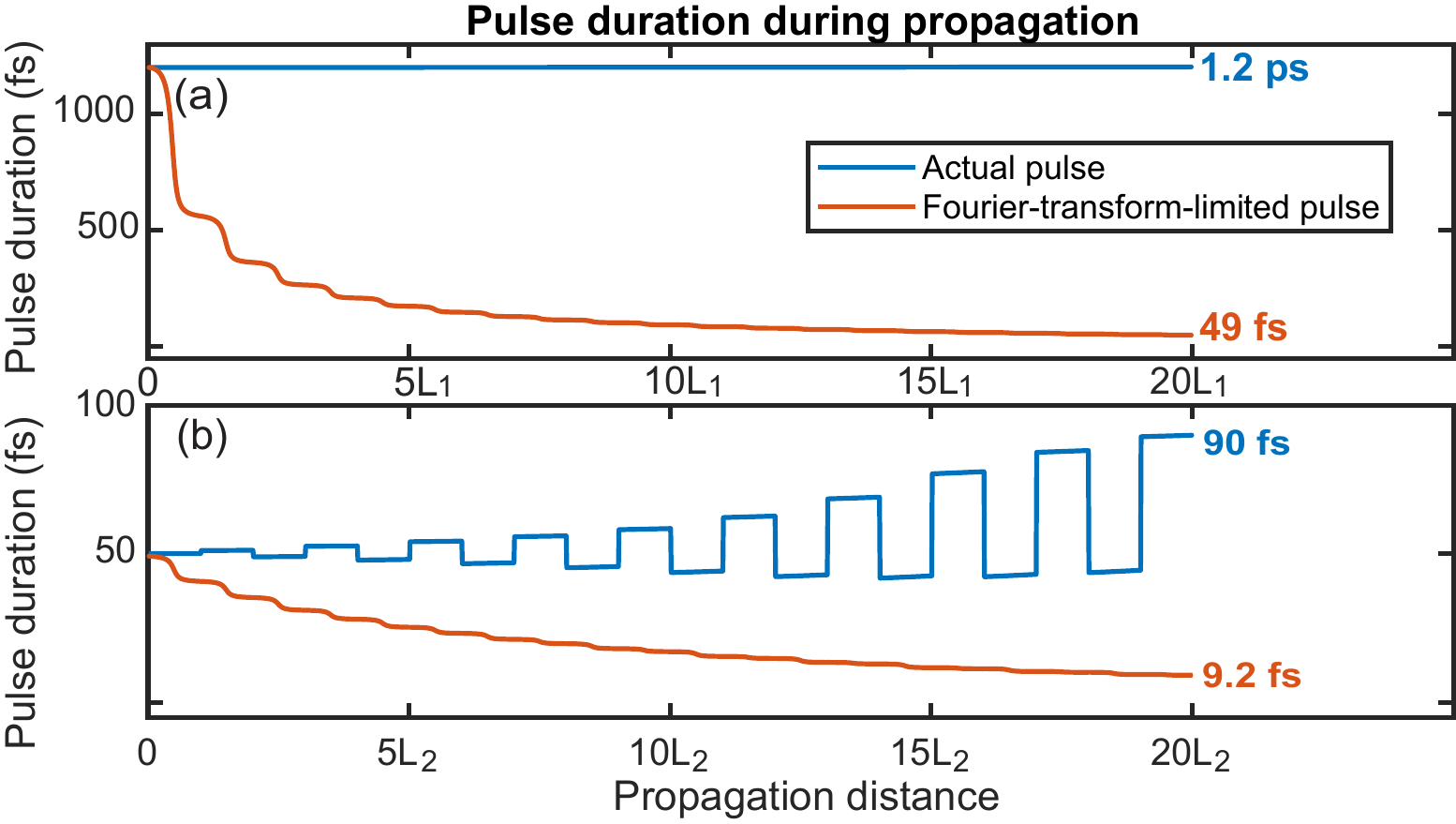}
\caption{Simulated post-compression using a two-stage cascaded MPC setup. Panels (a) and (b) show the pulse duration and corresponding FTL during propagation in the first and second cell in units of cell lengths $L_{1,2}$, respectively.}
\label{fig:concept}
\end{figure}

Our concept is demonstrated first using simulations shown in Fig.~\ref{fig:concept}. A 1.2-ps Gaussian pulse is compressed down to 9\,fs using two cascaded gas-filled MPCs (denoted as MPC1 and MPC2) each of them followed by a compressor for GDD removal. Low-dispersion quarter-wave-stack mirrors are used in the first stage, while a matched pair of dielectric mirrors with dispersion centered approximately at zero is used in the second stage. If linear dispersion plays a negligible role, neither the length of the cell nor the number of round trips inside the cell solely dictate the amount of B-integral (considering $B\ll 2\pi $ per pass). Consequently the spectral broadening characteristics can be adjusted via gas type and pressure to change the nonlinearity n$_2$ as desired. Hence, arbitrary cell lengths $L_1$ and $L_2$ are used in our simulations, considering 10 round trips for each cell.
The initial pulse duration in the first cell is roughly maintained during propagation within MPC1, while the Fourier-transform-limited (FTL) duration decreases. In contrast, the pulse duration in the second cell fluctuates due to the difference in the coatings of the mirror pair. The fluctuations get even stronger as the spectrum becomes broader, since the pulse becomes more sensitive to spectral phase variations. Although they do not hinder the spectral broadening process as the FTL of the broadened spectrum is still decreasing, it is important to take these fluctuations into account to prevent ionization in the nonlinear medium. 

\begin{figure}[!b]
\centering
\includegraphics[width=\linewidth]{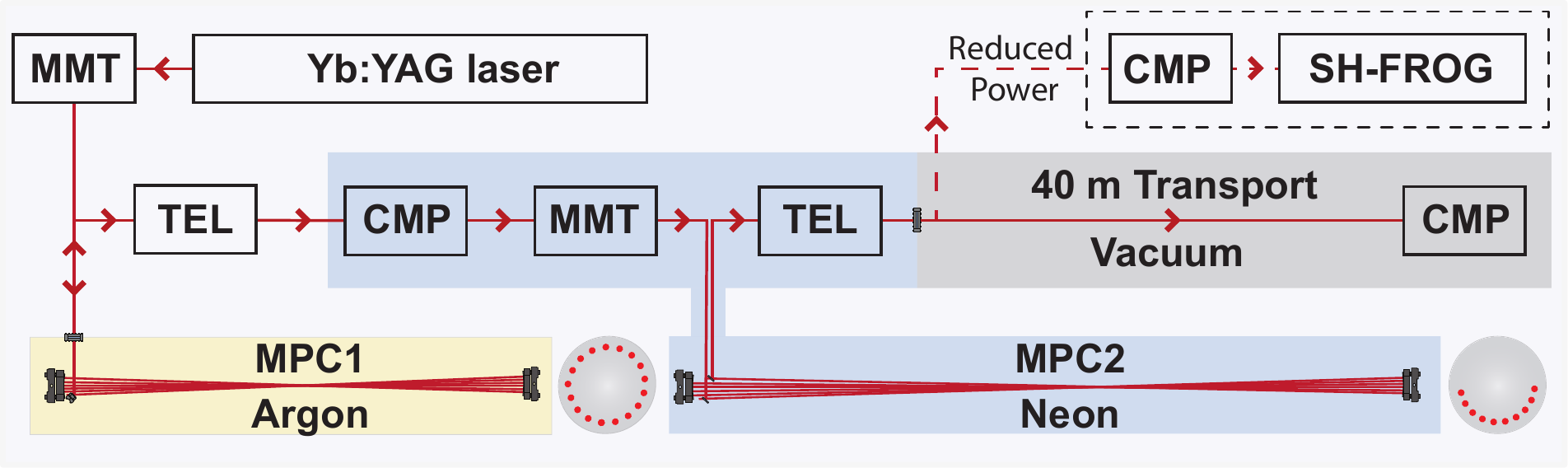}
\caption{Experimental setup used for post-compression comprising two gas-filled MPC stages (MMT: Mode-matching telescope, TEL: Telescope, CMP: Compressor), a pulse diagnostics section and a 40\,m long beam, transport section to an application area. The beam patterns obtained at the MPC mirrors are indicated schematically for both cells.}
\label{fig:Setup}
\end{figure}

We implement a similar post-compression concept employing a two-stage gas-filled MPC scheme in our experimental setup, as illustrated in Fig.~\ref{fig:Setup}. The 1.2\,ps pulses centered at 1030\,nm delivered from an Yb:YAG InnoSlab laser system have a pulse energy of 9.45\,mJ. For our experiments, the system is operated at 1\,kHz repetition rate. The laser beam is mode-matched to the eigenmode of MPC1 using a dielectric-mirror-based Galilean telescope. The first cell is 1.98\,m long, using two quarter-wave-stack focusing mirrors with a radius of curvature (ROC) of 1\,m \cite{rajhans_2021}.  
MPC1 is filled with 1200\,mbar of Argon as nonlinear medium to spectrally broaden the pulse resulting in a Fourier transform limit (FTL) duration of \textasciitilde50\,fs after 17 round trip through the cell. The throughput of the first cell is \textgreater94\,\% and the fluence expected on the mirrors calculated assuming linear mode-matching is about 180\,mJ/cm$^2$ which is almost a factor of five below the measured LIDT of the employed mirrors. Due to Kerr lensing at higher energy, nonlinear mode matching is required to keep the fluence on the mirrors below the damage threshold. A scraper mirror is used to direct the beam into the cell and to out-couple the beam at a slightly different angle in the vertical axis. The output beam is collimated and enlarged using another mirror-based telescope after passing through a standard variable attenuator consisting of an achromatic half-wave plate and a thin-film polarizer. The beam is then sent through a single-pass transmission grating compressor placed inside a vacuum chamber, providing a compressed output pulse duration close to the FTL. A single-pass configuration introduces an inevitable spatial chirp which could be kept down to a negligible amount due to a large beam size of ~15\,mm and a small grating distance of about 2\,mm. The single-pass configuration is compact, easy to adjust and enables a high throughput of 96\,\% providing an efficient alternative to chirped mirror-based compressors. The output of the compressor is sent via another mode matching telescope to the second MPC stage. MPC2 is placed in a vacuum chamber connected to the compressor chamber of MPC1 without windows, i.e. both chambers are filled with the gas used for MPC2.

\begin{figure}[t!]
\centering
\includegraphics[width=\linewidth]{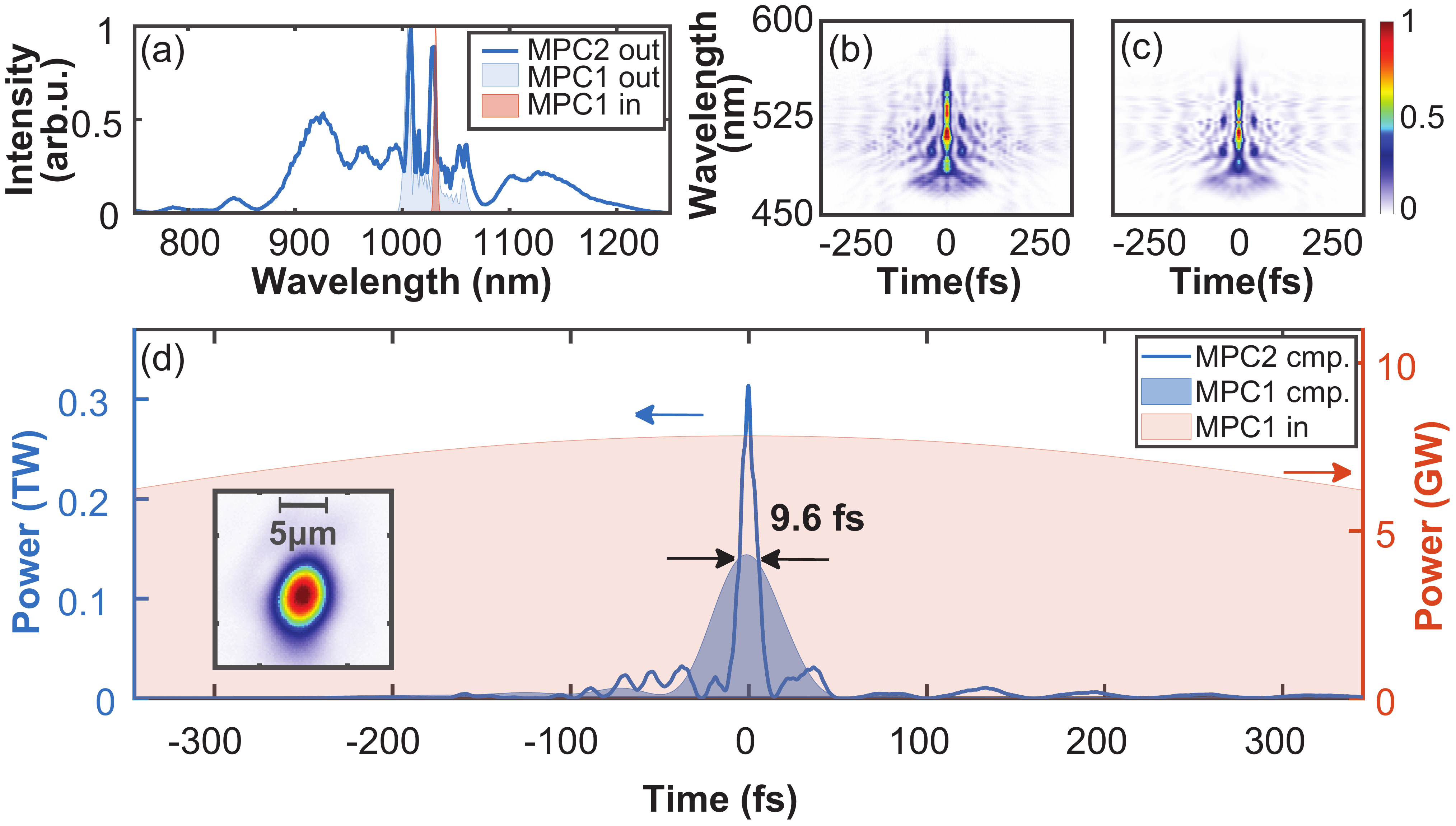}
\caption{(a) Measured spectra at input and output of MPC1 and output of MPC2. (b) and (c) show recorded and retrieved FROG traces of the MPC2 compressed output respectively. (d) MPC1 and MPC2 compressed output pulses retrieved using FROG. The figure also highlights the peak power enhancement from MPC1 input to MPC2 compressed output. Inset: Beam profile after focusing the compressed output of MPC2 reaching a spot size of  4.5 $\mu$m$\times$3.9 $\mu$m.}
\label{fig:Results}
\end{figure}

MPC2 employs a dispersion matched pair of focusing mirrors centered at zero GDD, with an ROC of 1.74\,m.
Compared to quarter-wave stack dielectric mirrors, broadband dispersion-optimized mirrors typically support only a much reduced LIDT. MPC2 thus needs to be longer than MPC1 in order to reduce the fluence at the mirrors. In-house LIDT characterization of the MPC2 mirrors yielded a damage threshold of 150\,mJ/cm$^2$.  
In order to further reduce the fluence while optimizing throughput, MPC2 is configured for 20 round trips, however, the beam is coupled out after the 10th round trip, i.e. only half of a circular multi-pass pattern is formed on the MPC mirrors. An MPC designed for a larger number of round trips is beneficial for keeping the fluence on the mirrors low. The relationship between the fluence $F$ on the mirrors and the number of round trips $N$ can be approximated for a standard Herriott-type MPC operated close to the stability edge as \cite{Viotti_MPC_review}:
\begin{equation}
    F \approx \frac{\pi E}{R\lambda N},
    \label{eq:fluence}
\end{equation}
where $E$ is the pulse energy, $R$ is the radius of curvature of the mirrors, and $\lambda$ the central wavelength. Here, $N$ is limited by the mirror size and efficient separation of input and output beams. Taking the beam out exactly halfway through the Herriott cell ensures that the Guoy phase accumulated is $\pi$ \cite{Viotti_MPC_review} and an input-to-output imaging condition is still satisfied, benefiting beam pointing characteristics. The fluence on the mirrors is about 80\,mJ/cm$^2$ for this configuration, about 1.9 times below the measured LIDT. The nonlinear gaseous medium is chosen such that the peak intensity at the focus does not result in ionization. Within 10 round trips through the cell, sufficient B-integral can be accumulated with 620\,mbar of neon. The MPC2 throughput is 88\,\% yielding an output spectrum corresponding to a FTL of \textasciitilde7\,fs, as shown in Fig.~\ref{fig:Results}(a). The spectrally broadened output is sent through a pair of fused silica wedges (5 degree wedge angle, with anti-reflection coating) to gain fine dispersion control, and the output pulse energy after the wedges is 7.1\,mJ. At this level of broadening, the third-order dispersion (TOD) from fused silica cannot be neglected. Thus, we use specifically designed chirped mirror pairs matched to the dispersion of fused silica, compensating 6\,mm fused silica per double reflection. In addition, all subsequent chamber windows are made of fused silica. 

The attenuated compressed pulses are sent to a home-built second harmonic frequency-resolved optical gating (SH-FROG) setup for characterization, yielding a duration of 9.6\,fs as shown in Fig.~\ref{fig:Results}(d). Recorded and retrieved FROG traces are provided in Fig.~\ref{fig:Results}(b) and (c), respectively. 
The MPC2 output is then transported towards an application area using 2-inch dielectric matched-pair mirrors placed in vacuum. Following beam transport, the pulses are compressed at full power using a second compressor made of 2-inch compressor mirrors.  All windows have anti-reflection coatings to keep losses at minimum, and are also made as thin as possible within safety standards to minimize B-integral. The throughput of the compressor is measured to be 95\,\% resulting in a high output peak power of about 0.31\,TW, corresponding to a peak power boost of about 40.
After around 40\,m of beam transport, the compressed pulses are focused down to a few $\mu$m spot using an off-axis parabola. The corresponding beam profile is shown in the inset of Fig.~\ref{fig:Results}(d). 

\begin{figure}[!th]
\centering
\includegraphics[width=\linewidth]{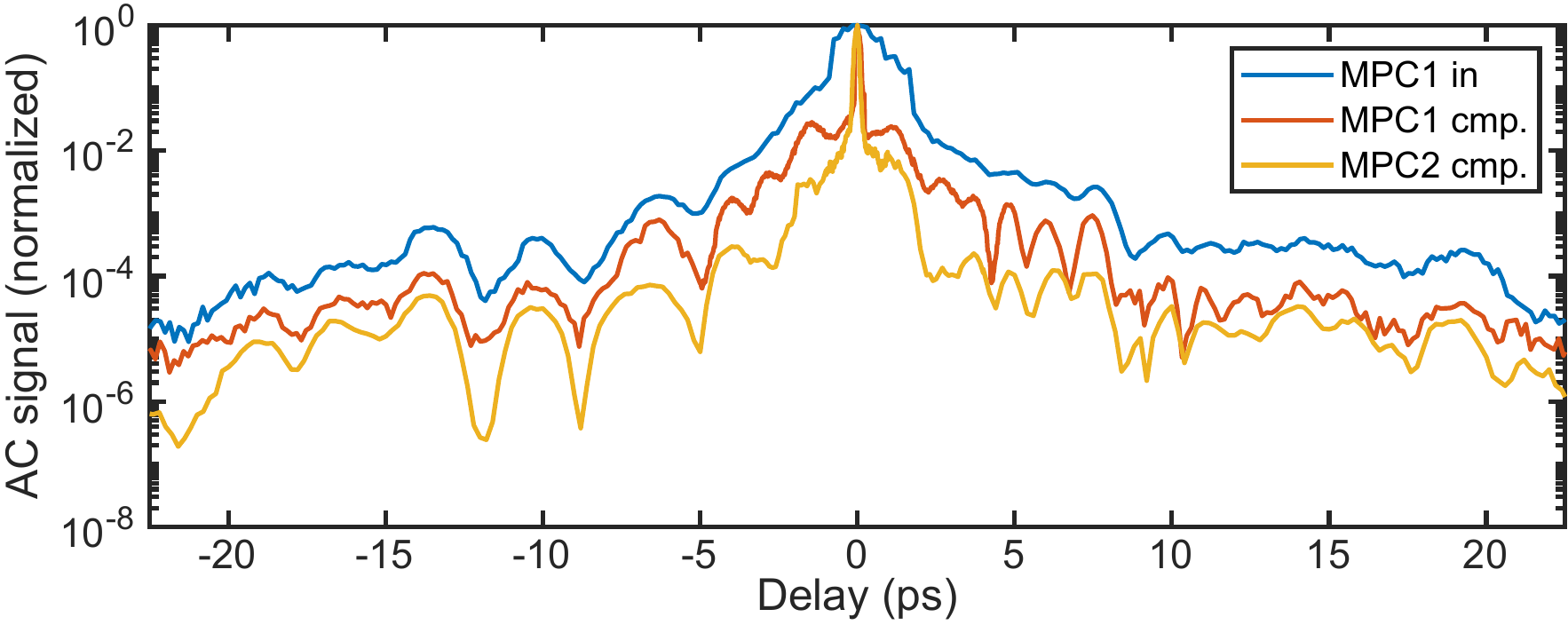}
\caption{Third-order autocorrelation (AC) traces measured at the output of the laser and after first and second compression stage.}
\label{fig:3rdorder}
\end{figure}

We further experimentally address the demand for high-peak power laser pulses comprising high temporal contrast by fully characterizing input and output pulses of both post-compression stages using a commercial third-order autocorrelator (AC). Fig.~\ref{fig:3rdorder} displays the recorded AC traces, showing  an improvement in temporal contrast by about one order of magnitude at picosecond delays. The temporal contrast improves after each post-compression stage as only the main part of the pulse experiences peak power enhancement while the weaker temporal pedestals remain unaffected. The peak power enhancement is larger in the first stage and thus, the contrast improvement is also greater. It should be noted that the instrument-response function of the employed third-order AC likely prevents to fully temporally resolve the post-compressed pulses after MPC1 and MPC2, leading to an underestimated temporal contrast boost. 

In summary, we have demonstrated efficient post-compression of a multi-millijoule picosecond Yb laser system to sub-10\,fs approaching the terawatt regime. This was achieved using a dual-stage MPC setup employing dielectric mirrors. Despite the large compression factor exceeding 120, a compressed pulse of high temporal quality was obtained with a temporal contrast exceeding the input pulse contrast by at least one order of magnitude.  

High peak and average power laser sources will prospectively play an important role in advancing areas like laser driven particle sources, for applications such as ultrafast electron diffraction and high dose rate radiobiology \cite{Albert_2021_roadmap}. In addition, other fields employing laser-driven photon sources such as attosecond science or nanolithography may greatly benefit from multi-kHz TW-class lasers. The energy scaling efforts demonstrated in this work can potentially be advanced further. While Yb-lasers at higher average power and pulse energy have been demonstrated already \cite{Herkommer_Yb_system_2020,Wandt_2020}, MPC-based post-compression demands alternative cell geometries to support much higher pulse energies while ensuring a compact setup footprint. These may include e.g. bow tie MPCs \cite{Heyl_2022_bow_tie}, convex-concave MPCs \cite{Hariton_convex_concave_2023,Omar_convex_concave_2023}, or multiplexing approaches \cite{Stark_2022_temporal_stacking_MPC}, promising compact high repetition rate TW-scale lasers. 
\par


\subsubsection*{Funding and Acknowledgments} We acknowledge DESY (Hamburg, Germany), a member of the Helmholtz Association HGF, for support and the provision of experimental facilities. 

\subsubsection*{Disclosures} The authors declare no conflicts of interest.

\subsubsection*{Data availability} Data underlying the results presented in this paper are not publicly available at this time, however may be obtained from the authors upon reasonable request.

\subsubsection*{$^\dagger$These authors contributed equally to this work}


\bibliography{mybibliography}


\end{document}